\documentclass[art10]{article}
\usepackage{euscript,amssymb}
\def\stackunder#1#2{\mathrel{\mathop{#2}\limits_{#1}}}

\def\stackunder#1#2{\mathrel{\mathop{#2}\limits_{#1}}}
\catcode`\@=11

\newcommand{\be}[1]{\begin{equation}\label{#1}}
\newcommand{\ee}{\end{equation}}
\newcommand{\ba}[1]{\begin{eqnarray}\label{#1}}
\newcommand{\ea}{\end{eqnarray}}
\newcommand{\rf}[1]{(\ref{#1})}
\newcommand{\nn}{\nonumber}
\newcommand{\ov}{\overline}
\newcommand{\td}{\tilde}
\newcommand{\bmatrix}[1]{\left( \begin{array}{#1}}
\newcommand{\ematrix}{\end{array}\right)}
\newcommand{\opensquare}{\mbox{$\rlap{$\sqcap$}\sqcup$}}

\newcommand{\sign}{ \mbox{\rm sign}\,}

\title{Asymptotical AdS from non-linear gravitational models
with stabilized extra dimensions}
\author{U. G\"unther$^a$\footnote{e-mail: u.guenther@fz-rossendorf.de}~\footnote{present address:
 Research Center Rossendorf, P.O. Box 510119, 01314 Dresden, Germany}\, ,
P. Moniz$^b$\footnote{e-mail: pmoniz@dfisica.ubi.pt}~
\footnote{Also at Centra --- IST, Rua Rovisco Pais, 1049 Lisboa,
Portugal} \, and A.
Zhuk$^c$ \footnote{e-mail: zhuk@paco.net}\\[2ex] $^a$
Gravitationsprojekt, Mathematische Physik I,\\
Institut f\"ur Mathematik,
Universit\"at Potsdam,\\
Am Neuen Palais 10, PF 601553, D-14415 Potsdam, Germany
\\[1ex] $^b$ Departamento de F$\acute{\i}$sica,
Universidade da Beira Interior,\\ Rua Marqu$\hat{e}$s
D'$\acute{A}$vila e Bolama, 6200 Covilh$\tilde{a}$, Portugal
\\[1ex] $^c$ Department of Physics, University of Odessa,\\ 2
Dvoryanskaya St., Odessa 65100, Ukraine }

\date{29.05.2002}
\begin{document}
\maketitle
\begin{abstract}
We consider non-linear gravitational models with a
multidimensional warped product geometry. Particular attention is
payed to  models with quadratic scalar curvature terms. It is
shown that for certain parameter ranges, the extra dimensions are
stabilized if the internal spaces have negative constant
curvature. In this case, the 4--dimensional effective cosmological
constant as well as the bulk cosmological constant become
negative. As a consequence, the homogeneous and isotropic external
space is asymptotically $\mbox{AdS}_4$. The connection between the
D--dimensional and the 4--dimensional fundamental mass scales sets
a  restriction on the parameters of the considered non-linear
models.
\end{abstract}

\section{\label{intro}Introduction}


Multidimensionality of our Universe is one of the most intriguing
assumption in modern physics. It follows naturally from theories
unifying different fundamental interactions with gravity, e.g.
M/string theory \cite{pol-wit}. The idea has received a great deal
of renewed attention over the last few years within the
"brane-world" description of the Universe. In this approach the
$SU(3)\times SU(2)\times U(1)$ standard model (SM) fields are
localized on a $3-$dimensional space-like hypersurface (brane)
whereas the gravitational field propagates in the whole (bulk)
space-time. The framework also implies that usual $4-$dimensional
physics is located on the brane (i.e. our Universe). Moreover,
brane-world physics provides a possible solution of the hierarchy
problem due to the well known connection between the Planck scale
$M_{Pl(4)}$ and the fundamental scale $M_{*(4+D^{\prime})}$ of
the $4-$dimensional and the (4+$D^{\prime}$)-dimensional
space-time, respectively:
\be{0.1} M_{Pl(4)}^2 \sim V_{D^{\prime}}
M_{*(4+D^{\prime})}^{2+D^{\prime}}\, . \ee
Here $V_{D^{\prime}}$ denotes the volume of the compactified
$D^{\prime}$ extra dimensions. It was realized in
\cite{sub-mill1,sub-mill1a,sub-mill2} that the localization of the
SM fields on the brane allows to lower $M_{*(4+D^{\prime})}$ down
to the electroweak scale $M_{EW} \sim 1$TeV without contradiction
with present observations. Therefore, the compactification scale
of the internal space can be of order
\be{0.2} r \sim V_{D^{\prime}}^{1/D^{\prime}} \sim
10^{\frac{32}{D^{\prime}}-17} \mbox{cm}\, . \ee
In this ADD model \cite{sub-mill1}, physically acceptable values
correspond to $D^{\prime}\ge 3$ (see e.g. \cite{experiment1}), and
for $D^{\prime} =3$ one arrives at a sub-millimeter
compactification scale $r\sim 10^{-6} \mbox{cm}$ of the internal
space.  Additionally, the geometry is assumed to be factorizable
as in the standard Kaluza-Klein (KK) model.  I.e., the topology is
the direct product of a non-warped external space-time manifold
and internal space manifolds with warp factors which depend on the
external coordinates. Beside this, the M-theory inspired
RS-scenario \cite{RS} represents an interesting approach with
non-factorizable geometry and $D^{\prime} =1$. Here, the
$4-$dimensional space-time is warped with a factor $\tilde\Omega$
which depends on the extra dimension and equation \rf{0.1} is
modified as follows: $M_{Pl(4)} \sim \tilde\Omega^{-1}M_{EW}$. In
our paper we shall  concentrate on the factorizable geometry of
the ADD-model.

According to observations the internal space should be static or
nearly static at least from the time of primordial
nucleosynthesis, (otherwise the fundamental physical constants
would vary). This means that at the present evolutionary stage of
the Universe  the compactification scale of the internal space
should either be stabilized and trapped at the minimum of some
effective potential, or it should be slowly varying (similar to
the slowly varying cosmological constant in the quintessence
scenario \cite{WCOS}). In both cases, small fluctuations over
stabilized or slowly varying compactification scales (conformal
scales/geometrical moduli) are possible.

Stabilization of extra dimensions (moduli stabilization) in models
with large extra dimensions (ADD models) has been considered in a
number of papers (see e.g., Refs.
\cite{sub-mill2,d2,sub-mill3,CGHW,Geddes,demir,NSST,PS})\footnote{In
most of these papers, moduli stabilization was  considered without
regard to the energy-momentum localized on the brane. A brane
matter contribution was taken into account, e.g., in \cite{PS}.}. In
the corresponding  approaches, a product topology of the
$(4+D^{\prime })-$dimensional bulk space-time was constructed from
Einstein spaces with scale (warp) factors depending only on the
coordinates of the external $4-$dimensional component. As a
consequence, the conformal excitations have the form of massive
scalar fields living in the external space-time. Within the
framework of multidimensional cosmological models (MCM)
 such excitations were investigated in \cite{GZ1,GZ,GZ(PRD2)} where
 they were called
 gravitational excitons. Later, since the ADD
compactification approach these geometrical moduli excitations are
known as radions \cite{sub-mill2,sub-mill3}. It should be noted
that over the last years the term radion has been used to describe
quite different forms of metric perturbations within brane-world
models. In MCM with warped product topology of the internal spaces
they are understood as conformal excitations of the additional
dimensions (gravitational excitons), whereas in RS-I-type models
they describe the relative motion of branes \cite{CGR}\footnote{A
 detailed discussion of radion stabilization and dynamics in RS models is given, e.g., in \cite{Csaki1,Csaki2}.
 An extended list of references on this topic can be found in \cite{MBLSZ}.}. The
differences between these two frameworks have been pointed out in
 \cite{BDL,CF}.

All above mentioned papers are devoted to the stabilization of
large extra dimension in theories with linear multidimensional
gravitational action. String theory suggests that the usual linear
Einstein-Hilbert action should be extended with higher order
non-linear curvature terms. In the present paper we use a
simplified approach with multidimensional Lagrangian of the form
$L = f(R)$, where  $f(R)$ is an arbitrary smooth function of the
 scalar curvature. Without connection to stabilization of the
extra-dimensions, such models ($4-$dimensional as well as
multi-dimensional ones) were considered  e.g. in Refs.
\cite{Kerner}. There, it was shown that the non-linear models are
equivalent to models with linear gravitational action plus a
minimally coupled scalar field with self-interaction potential.

In the present paper we advance this equivalence towards
investigating the problem of extra dimensions stabilization.
 We find that the stabilization of extra
dimensions takes place only if additional internal spaces have a
compact hyperbolic geometry and the effective $4-$dimensional
cosmological constant is negative. If the external space $M_0$ is
homogeneous and isotropic this implies that $M_0$ becomes
asymptotically anti-deSitter ($\mbox{AdS}_{D_0}$).
 Additionally, we show that
requiring  the extra dimensions to be dynamically stabilized is a
sufficient condition for the bulk space-time to acquire a constant
negative curvature.

The paper is structured as follows. After explaining the general
setup of our model in section \ref{setup}, we concretize the geometry to a
warped product of $n$ internal spaces. We perform a dimensional
reduction of the action functional to a 4-dimensional effective
theory with $(n+1)$ self-interacting minimally coupled scalar
fields (section \ref{reduction}). The stabilization of the extra dimensions is
 then reduced to the condition that the obtained
 effective potential for these fields should have a minimum.
In section \ref{stab}, a detailed analysis of this problem is given  for a
model with one internal space. The main results are summarized and
discussed in the concluding section \ref{conclu}.


\section{\label{setup}General theory}


We consider a $D= (4+D^{\prime})$ -- dimensional non-linear pure
gravitational theory with action
\be{1.1} S = \frac {1}{2\kappa^2_D}\int_M d^Dx \sqrt{|\ov g|}
f(\ov R)\; , \ee
where $f(\ov R)$ is an arbitrary smooth function with mass
dimension $\mathcal{O}(m^2)$ \ ($m$ has the unit of mass) of a
scalar curvature $\ov R = R[\ov g]$ constructed from the
D--dimensional metric $\ov g_{ab}\; (a,b = 1,\ldots,D)$.
\be{1.1a}  \kappa^2_D = 8\pi /
M_{*(4+D^{\prime})}^{2+D^{\prime}}\ee
is the D--dimensional gravitational constant (subsequently, we
 assume that $M_{*(4+D^{\prime})}\sim M_{EW}$). The equation
of motion for this theory reads  \cite{Kerner}
\be{1.2} f^{\prime }\ov R_{ab} -\frac12 f\, \ov g_{ab} - \ov
\nabla_a \ov \nabla_b f^{\prime } + \ov g_{ab} \ov{\opensquare }
f^{\prime } = 0\; , \ee
where $ f^{\prime } =df/d\ov R$, $\; \ov R_{ab} = R_{ab}[\ov g]$.
\ $\; \ov \nabla_a$ is the covariant derivative with respect to
the metric $\ov g_{ab}$; and the corresponding Laplacian is
denoted by
\be{1.3} \ov{\opensquare} = \opensquare [\ov g] = \ov g^{ab}\ov
\nabla_a \ov \nabla_b = \frac{1}{\sqrt{|\ov g|}}
\partial_a \left( \sqrt{|\ov g|}\quad \ov g^{ab}
\partial_b \right)\; .
\ee
Eq. \rf{1.2} can be rewritten in the form
\be{1.4} f^{\prime }\ov G_{ab} +\frac12 \ov g_{ab} \left( \ov R
f^{\prime} - f\right) - \ov \nabla_a \ov \nabla_b f^{\prime } +
\ov g_{ab} \ov{\opensquare } f^{\prime } = 0\; , \ee
where $\ov G_{ab} = \ov R_{ab} -\frac12 \ov R \; \ov g_{ab}$. The
trace of eq. \rf{1.2} is
\be{1.5} (D-1)\ov{\opensquare } f^{\prime } = \frac{D}{2} f
-f^{\prime }\ov R\;  \ee
and can be considered as a connection between $\ov R$ and $f$.

It is well known, that for $f'(\ov R) >0$  the conformal
transformation
\be{1.6} g_{ab} = \Omega^2 \ov g_{ab}\; , \ee
with
\be{1.7} \Omega = \left[ f'(\ov R)\right]^{1/(D-2)}\;  ,\ee
reduces the non-linear theory \rf{1.1}  to a linear one with an
additional scalar field. The equivalence of the theories  can be
easily proven with the help of the following auxiliary formulas:
\be{1.8} \opensquare = \Omega^{-2}\left[\ov{\opensquare } +
(D-2)\ov g^{ab}\Omega^{-1}\Omega_{,a}\partial_b\right]
\Longleftrightarrow \ov {\opensquare} = \Omega^{2}\opensquare -
(D-2) g^{ab}\Omega \; \Omega_{,a}\partial_b \; , \ee
\be{1.9} R_{ab} = \ov R_{ab} +\frac{D-1}{D-2}( f')^{-2} \ov
{\nabla}_a f' \ov {\nabla}_b f' -(f')^{-1} \ov {\nabla}_a \ov
{\nabla}_b f'- \frac{1}{D-2} \ov g_{ab} (f')^{-1} \ov
{\opensquare} f' \ee
and
\be{1.10} R = (f')^{2/(2-D)}\left\{ \ov R +\frac{D-1}{D-2}
(f')^{-2} \ov g^{ab}\partial_a f'\partial_b f'-
2\frac{D-1}{D-2}(f')^{-1} \ov {\opensquare} f'\right\}\; . \ee
Thus, eqs. \rf{1.4} and \rf{1.5} can be rewritten
as
\be{1.11} G_{ab} = \phi_{,a}\phi_{,b} -\frac12
g_{ab}g^{mn}\phi_{,m}\phi_{,n} - \frac12 g_{ab}\; e^{\frac
{-D}{\sqrt{(D-2)(D-1)}}\phi} \left(\ov R f'- f\right) \ee
and
\be{1.12} \opensquare \phi = \frac {1}{\sqrt{(D-2)(D-1)}}\;
e^{\frac {-D}{\sqrt{(D-2)(D-1)}}\phi} \left(\frac {D}{2} f-f'\ov R
\right)\; , \ee
where \be{1.13} f' = \frac {df}{d \ov R} := e^{A \phi} > 0\;
,\quad A := \sqrt{\frac{D-2}{D-1}}\; . \ee
Eq. \rf{1.13} can be used to express $\ov R$ as a function of the
dimensionless field $\phi$ : $\ov R = \ov R( \phi )$.

It is easily seen that eqs. \rf{1.11} and \rf{1.12} are the
equations of motion for the action
\be{1.14} S = \frac{1}{2\kappa^2_D} \int_M d^D x \sqrt{|g|} \left(
R[g] - g^{ab} \phi_{,a} \phi_{,b} - 2 U(\phi )\right)\; , \ee
where
\be{1.15}
U(\phi ) = \frac12 e^{- B \phi}
\left[\; \ov R (\phi )e^{A \phi } - f\left( \ov R (\phi )\right) \right]\; ,
\quad B := \frac {D}{\sqrt{(D-2)(D-1)}}
\ee
and they can be written as follows:
\be{1.16} G_{ab} = T_{ab}\left[ \phi , g \right]\; , \ee
\be{1.17}
\opensquare \phi = \frac {\partial U(\phi)}{\partial \phi}\; .
\ee
Here, $T_{ab}\left[ \phi , g \right]$ is the standard expression
of the energy--momentum tensor for the minimally coupled scalar
field with potential \rf{1.15}. Eq. \rf{1.17} can be considered as
a constraint equation following from the reduction of the
non-linear theory \rf{1.1} to the linear one \rf{1.14}.

Let us consider what will happen if, in some  way, the scalar
field $\phi$ tends asymptotically to a constant: $\phi \to
\phi_{0} $. From eq. \rf{1.13} we see that in this limit the
non-linearity disappears and  \rf{1.1} becomes a linear theory
 $f(\ov R) \sim c_1 \ov R + c_2$ with $c_1 = f' = \exp(A
\phi_{0})$ and a cosmological constant $-c_2/(2c_1)$. In the case
of homogeneous and isotropic space-time manifolds, linear purely
geometrical theories with constant $\Lambda -$term necessarily
imply an (A)dS geometry. Thus, in the limit $\phi \to \phi_{0}$
the D--dimensional theory \rf{1.1} can asymptotically lead to an
(A)dS with scalar curvature:
\be{1.17a} \ov R \rightarrow -\frac{D}{D-2}\, \frac{c_2}{c_1}\, .
\ee
Clearly, the linear theory \rf{1.14} would reproduce this
asymptotic (A)dS-limit for $\phi \to \phi_{0}$:
\be{1.17b} R \rightarrow  2\frac{D}{D-2}\, U(\phi_0) = -
\frac{D}{D-2}\; c_2\,  c_1^{-\frac{D}{D-2}}\, . \ee
Hence, in this limit $\ov R / R \to c_1^{\frac{D}{D-2}}$ in
accordance with eq. \rf{1.10} and $f'= c_1$. In section \ref{stab} we shall
show that the stabilization of the extra dimensions automatically
results in condition $\phi \to \phi_{0} $ with $U(\phi_0) < 0$.
Thus, the D-dimensional space-time (bulk) can become asymptotically
$\mbox{AdS}_D$.

In the rest of the paper we consider the quadratic theory:
\be{1.18} f(\ov R ) = \ov R + \alpha \ov R^{\; 2} - 2\Lambda _D\;
,\ee
where the parameter $\alpha$ has dimensions $\mathcal{O}(m^{-2})$.
 For this theory we obtain
\be{1.19} 1 + 2\alpha \ov R = e^{A \phi} \Longleftrightarrow \ov R
= \frac{1}{2\alpha } \left( e^{A \phi } - 1\right) \ee
and
\be{1.20} U(\phi ) = \frac12 e^{-B \phi }\left[ \frac{1}{4\alpha
}\left( e^{A \phi } - 1\right)^2 + 2 \Lambda _D \right]\; . \ee
The condition $f' > 0$ implies $1+2\alpha \ov R
>0$.



\section{\label{reduction}Dimensional reduction}

In this section we assume that the D-dimensional bulk space-time
$M$ undergoes a spontaneous compactification  to a warped product
 manifold
\be{2.1} M = M_0 \times M_1 \times \ldots \times M_n \ee
with  metric
\be{2.2} \ov g=\ov g_{ab}(X)dX^a\otimes dX^b=\ov
g^{(0)}+\sum_{i=1}^ne^{2\ov {\beta} ^i(x)}g^{(i)}\; . \ee
The coordinates on the $(D_0=d_0+1)$ - dimensional manifold $M_0 $
(usually interpreted as our $(D_0=4)$-dimensional Universe) are
denoted by $x$ and the corresponding metric by
\be{2.3} \ov g^{(0)}=\ov g_{\mu \nu }^{(0)}(x)dx^\mu \otimes
dx^\nu\; . \ee
Let the internal factor manifolds $M_i$ be $d_i$-dimensional
Einstein spaces with metric
$g^{(i)}=g^{(i)}_{m_in_i}(y_i)dy_i^{m_i}\otimes dy_i^{n_i},$ i.e.,
\be{2.4} R_{m_in_i}\left[ g^{(i)}\right] =\lambda
^ig_{m_in_i}^{(i)},\qquad m_i,n_i=1,\ldots ,d_i \ee
and
\be{2.5} R\left[ g^{(i)}\right] =\lambda ^id_i\equiv R_i \sim
r_i^{-2}\; , \ee
where $r_i = \left( \int d^{d_i}y \sqrt{|g^{(i)}|}\right)^{1/d_i}$
is a characteristic size of $M_i$. For the metric ansatz \rf{2.2}
the scalar curvature $\ov R$ depends only on $x$: $\ov R[\ov g] =
\ov R(x)$. Thus $\phi$ is also a function of $x$ : $\phi = \phi
(x)$.

The conformally transformed metric \rf{1.6} reads
\be{2.6} g = \Omega^2 \ov g = \left( e^{A \phi }\right)^{2/(D-2)}
\ov g\: := g^{(0)}+\sum_{i=1}^ne^{2 \beta^i(x)}g^{(i)}\;  \ee
with
\be{2.7} g^{(0)}_{\mu \nu} = \left( e^{A \phi}\right)^{2/(D-2)}
\ov g^{(0)}_{\mu \nu}\; , \ee \be{2.8} \beta^i = \ov {\beta}^i +
\frac{A}{D-2} \phi\; . \ee
The fact that the fields $\phi $ and $\beta^i $ depend only on $x$
allows us to perform the dimensional reduction of action
\rf{1.14}. Without loss of generality we set the compactification
scales of the internal spaces at present time at $\beta^i = 0 \;
(i = 1,\ldots ,n)$. The corresponding total volume of the internal
spaces is given by
\be{2.7a} V_{D^{\prime }} \equiv
\prod_{i=1}^n\int\limits_{M_i}d^{d_i}y \sqrt{|g^{(i)}|} =
\prod_{i=1}^n r_i^{d_i}\; , \ee
where $V_{D^{\prime }}$ has dimensions $\mathcal{O}(m^{-D^{\prime}})$,
and $D^{\prime} = D-D_0 = \sum_{i=1}^n d_i$ is the number of the
extra dimensions. After  dimensional reduction action \rf{1.14}
reads
$$
S=\frac 1{2\kappa _0^2}\int\limits_{M_0}d^{D_0}x\sqrt{|g^{(0)}|}%
\prod_{i=1}^ne^{d_i\beta ^i}\left\{ R\left[ g^{(0)}\right] -G_{ij}g^{(0)\mu
\nu }\partial _\mu \beta ^i\,\partial _\nu \beta ^j
-g^{(0)\mu \nu}\partial_{\mu} \phi \partial_{\nu } \phi +\right.
$$
\be{2.9} +\sum_{i=1}^n\left. R\left[ g^{(i)}\right] e^{-2\beta
^i}-2 U(\phi) \right\} \; , \ee
where  $G_{ij}=d_i\delta _{ij}-d_id_j\ (i,j=1,\ldots ,n)$ is the
midisuperspace metric \cite{IMZ, RZ} and
\be{2.9a} \kappa^2_0 := \frac{\kappa^2_D}{V_{D^{\prime }}} \ee
is the $D_0-$dimensional (4-dimensional) gravitational constant.
If we take the electroweak scale $M_{EW}$ and the Planck scale
$M_{Pl}$ as fundamental ones for $D-$dimensional (see eq.
\rf{1.1a}) and 4-dimensional space-times ($\kappa^2_0 = 8\pi /
M^2_{Pl}$) respectively, then we reproduce eqs. \rf{0.1} and
\rf{0.2}.

Action \rf{2.9} is written in the Brans--Dicke frame.  Conformal
transformation to the Einstein frame \cite{GZ1,GZ}
\be{2.10} \td g_{\mu \nu }^{(0)}= {\left( \prod_{i=1}^ne^{d_i\beta
^i}\right) }^{\frac 2{D_0-2}}g_{\mu \nu }^{(0)} \ee
yields
\be{2.11} S=\frac 1{2\kappa
_0^2}\int\limits_{M_0}d^{D_0}x\sqrt{|\td g^{(0)}|}\left\{ R\left[
\td g^{(0)}\right] -\bar G_{ij}\td g^{(0)\mu \nu }\partial _\mu
\beta ^i\,\partial _\nu \beta ^j- \td g^{(0) \mu \nu}
\partial_{\mu}\phi \partial_{\nu} \phi -2U_{eff} (\beta ,\phi )
\right\} \; . \ee
The tensor components of the midisuperspace metric (target space
metric on $\mathbb{R}_T^n$ ) $\bar G_{ij}\ (i,j=1,\ldots ,n)$ ,
its inverse metric $\bar G^{ij}$ and the effective potential are
respectively \be{2.12} \bar G_{ij}=d_i\delta _{ij}+\frac
1{D_0-2}d_id_j\; , \ee \be{2.13} \bar G^{ij}=\frac{\delta
^{ij}}{d_i}+\frac 1{2-D} \ee and \be{2.14} U_{eff}(\beta ,\phi )
={\left( \prod_{i=1}^n e^{d_i\beta ^i}\right) }^{-\frac 2{D_0-2}}
\left[ -\frac 12\sum_{i=1}^nR_ie^{-2\beta ^i}+ U (\phi ) \right]
\; . \ee



\section{\label{stab}Stabilization of the internal space}

Without loss of generality\footnote{The only difference between a
general model with $n>1$ internal spaces and the particular one
with $n=1$ consists in an additional diagonalization of the
geometrical moduli excitations.}, we consider in the present
section a model with only one $d_1$-dimensional internal space.  The corresponding
action \rf{2.11} reads
\be{3.1} S=\frac 1{2\kappa
_0^2}\int\limits_{M_0}d^{D_0}x\sqrt{|\td g^{(0)}|}\left\{ R\left[
\td g^{(0)}\right] - \td g^{(0) \mu \nu}
\partial_{\mu}\varphi
\partial_{\nu} \varphi - \td g^{(0) \mu \nu} \partial_{\mu}\phi
\partial_{\nu} \phi -2U_{eff} (\varphi ,\phi ) \right\} \; , \ee
where
\be{3.2} \varphi := -\sqrt{\frac{d_1(D-2)}{D_0-2}}\beta^1 \ee
and
\be{3.3} U_{eff}(\varphi ,\phi ) = e^{2\varphi
\sqrt{\frac{d_1}{(D-2)(D_0-2)}}} \left[ -\frac12 R_1e^{2\varphi
\sqrt{\frac{D_0-2}{d_1(D-2)}}}+ U (\phi ) \right] \; . \ee
For simplicity we continue to work with dimensionless scalar
fields $\varphi ,\phi $ instead of passing to canonical ones
(modulo $8\pi $): $\tilde{\varphi} = \varphi \, M_{Pl}, \,
\tilde{\phi} = \phi \, M_{Pl} $ and $\tilde{U}_{eff} = M_{Pl}^2
U_{eff}$. The restoration of the correct dimensionality is
obvious.

The equations of motion for $\varphi$ and $\phi$ are respectively
\be{3.4} \widetilde {\opensquare} \varphi = \frac{\partial
U_{eff}}{\partial \varphi}\; , \ee
\be{3.5} \widetilde {\opensquare} \phi = \frac{\partial
U_{eff}}{\partial \phi}\; , \ee
where
\be{3.6} \frac{\partial U_{eff}}{\partial \varphi} = 2\:
\sqrt{\frac{d_1}{(D-2)(D_0-2)}}\; U_{eff} - R_1\:
\sqrt{\frac{D_0-2}{d_1(D-2)}}\; e^{2\varphi
\sqrt{\frac{D-2}{d_1(D_0-2)}}} \ee
and
\be{3.7} \frac{\partial U_{eff}}{\partial \phi}= e^{2\varphi
\sqrt{\frac{d_1}{(D-2)(D_0-2)}}}\quad \frac{\partial
U(\phi)}{\partial \phi}\; . \ee

In order to obtain a stable compactification of the internal
space, the potential $U_{eff}(\varphi ,\phi )$ should have a
minimum with respect to $\varphi$ and $\phi$. This is obvious with
respect to the field $\varphi$ because it is precisely the
stabilization of this field that we aim to achieve. It is also
clear that potential $U_{eff}(\varphi ,\phi )$ should have a
minimum with respect to $\phi$ because without stabilization of
$\phi$ the effective potential remains a dynamical function and
the extremum condition $\left. \partial U_{eff}/\partial \varphi
\right|_{\varphi = 0} =0$ is not satisfied (see eq. \rf{3.6}).
Furthermore, eq. \rf{3.7} shows that the extrema of the potentials
$U_{eff}(\varphi ,\phi )$ and $U(\phi)$ with respect to the field
$\phi$ coincide with each other. Thus, the stabilization of the
extra dimension takes place iff the field $\phi$ goes to the
minimum of the potential $U(\phi)$. According to the discussion in
section \ref{setup} (see eqs. \rf{1.17a} and \rf{1.17b}) this  results in an
asymptotically constant curvature space-time (for a non-zero
minimum of $U(\phi)$).

Let us now present a detailed analysis of the quadratic
gravitational theory \rf{1.18} with potential $U(\phi )$
\rf{1.20}. First, we shall investigate the range of parameters
which ensures a minimum of  $U(\phi )$.
The extremum condition
gives
\be{3.8}
\partial_{\phi} U = 0 \Longrightarrow
(2A-B)x^2 + 2(B-A)x -(q+1)B = 0 \; , \ee
where $ x := e^{A\phi} > 0$ and $q := 8\alpha \Lambda _D$. The
non-negative solution of this equation defines the position of the
extremum:
\ba{3.9} x_0 = e^{A\phi_0} & = & \frac{-(B-A) + \sqrt{(B-A)^2 +
(2A-B)(q+1)B}}
{2A-B} \nn \\
& = & \frac{-(B-A)+\sqrt{A^2 + (2A-B)Bq}}{2A-B}\; .
\ea
From the inequalities
\be{3.10}
B-A = \frac{2}{\sqrt{(D-2)(D-1)}} > 0
\ee
and
\be{3.11}
2A-B = \frac{D-4}{\sqrt{(D-2)(D-1)}} > 0 \quad \mbox{for}\; D>4
\ee
it follows that the parameter $q$ should be restricted to the
half-line
\be{3.12} q = 8\alpha \Lambda _D > -1\; . \ee
The case $q=-1$ corresponds to $\phi_0\to -\infty$ and is not
considered in the following.

The necessary condition for the existence of a minimum of the
potential $U(\phi)$
\be{3.13} \left.
\partial^2_{\phi \phi} U(\phi) \right|_{extr} =
\frac{A}{4\alpha}e^{(A-B)\phi_0}\left[(2A-B)e^{A\phi_0} + (B-A)
\right] = \frac{1}{4\alpha}\frac{1}{D-1} x_0^{-\frac{2}{D-2}}
\left[(D-4)x_0 +2\right] > 0 \ee
requires positive values of the parameter
$\alpha >0$.
From the explicit expression of $U(\phi)$ at the extremum
\be{3.14} \left. U(\phi)\right|_{extr} =
\frac{1}{8\alpha}x_0^{-\frac{D}{D-2}}\left[(x_0-1)^2 + q\right]\;
, \ee
 it is easy to see that  $\left. U\right|_{min} \ge 0$ for
$\Lambda _D \ge 0$ and $\left. U\right|_{min} < 0$ for $\Lambda _D
< 0$.  In the latter case we have $-1<8\alpha\Lambda _D<0$.

Let us show now that the total potential $U_{eff}(\varphi , \phi
)$ also has a global minimum in the case when $U(\phi )$ has a
negative minimum. To prove it, it is convenient to rewrite
potential \rf{3.3} as follows
\be{3.15}
U_{eff}(\varphi ,\phi )\,=\,
\stackunder{F}{\underbrace{e^{2\varphi \sqrt{\frac{d_1}{(D-2)(D_0-2)}}}}}
\, \stackunder{G}{
\underbrace{\left[
-\frac12R_1e^{2\varphi \sqrt{\frac{D_0-2}{d_1(D-2)}}} +
U(\phi )
\right]}
}\, .
\ee
The extremum condition gives
\ba{3.16}
\partial_{\varphi }U_{eff} =
\left(2\sqrt{\frac{d_1}{(D-2)(D_0-2)}}\; G + \partial_{\varphi }G\right)F=0
&\Longrightarrow &
\partial_{\varphi }G =-2\sqrt{\frac{d_1}{(D-2)(D_0-2)}}\; G\; ,\label{3.16.1}\\
\partial_{\phi }U_{eff}=F(\partial _{\phi }U)=0
&\Longrightarrow &
\partial _{\phi }U=0\; ,\label{3.16.2}
\ea
whereas the eigenvalues of the Hessian at the minimum should be
non-negative
\ba{3.17}
\partial^2_{\varphi \varphi } U_{eff} &=&
\left[\partial^2_{\varphi \varphi }G-
4\frac{d_1}{(D-2)(D_0-2)}G
\right]F>0\; ,  \label{3.17.1}\\
\partial^2_{\phi \phi}U_{eff} &=& F\partial^2_{\phi \phi}U>0
\quad \Longrightarrow \quad
\partial^2_{\phi \phi}U>0\; ,\label{3.17.2}\\
\partial^2_{\varphi \phi }U_{eff}&=&
2\sqrt{\frac{d_1}{(D-2)(D_0-2)}}\; F \partial_{\phi }U=0\; .
\label{3.17.3} \ea
Choosing the compactification scale of the extra dimension at
$\beta^1_{min} = \varphi_{min} = 0$, we find the following
relations at the extremum
 \ba{3.19} R_1
&=&\frac{2d_1}{D-2} \left. U (\phi)\right|_{extr}\; ,\label{3.19.1}\\
\left.G\right|_{extr}&=&\frac{D_0-2}{D-2} \left. U (\phi)
\right|_{extr}\label{3.19.2} \ea
and hence
\be{3.20} \sign (R_1)=\sign (\left. U (\phi)\right|_{extr})= \sign
(\left.G\right|_{extr})\, . \ee
Using the obvious relation
\be{3.21}
\partial^2_{\varphi \varphi }G =
-2\frac{D_0-2}{d_1(D-2)} R_1e^{2\varphi
\sqrt{\frac{D_0-2}{d_1(D-2)}}} \ee
and eqs. \rf{3.17.1}, \rf{3.19.1},  \rf{3.19.2} we see that
\be{3.22} -\frac{4}{D-2}\left.U (\phi) \right|_{min}>0
\Longrightarrow \left.U (\phi)\right|_{min}<0 . \ee
This inequality sets strong restrictions on the considered non-linear
model:
\begin{enumerate}
\item According to eq. \rf{1.17b} it implies that
 the
stabilization of the extra dimension  leads asymptotically to a
negative constant curvature bulk space-time.
\item Only models with parameters from the range
$\alpha
> 0$ and $-1<8\alpha \Lambda _D <0$ will stabilize (see eqs.
\rf{3.13} and \rf{3.14}).
 All other configurations are
excluded.
\item  The global minimum of the whole effective potential
$U_{eff}$ is also negative:
\be{3.25} \left. U_{eff}\right|_{min} =
\frac{D_0-2}{D-2}
\left. U (\phi) \right|_{min} = \frac{D_0-2}{2d_1}\; R_1 \, <\,
0\; .
\ee
Its value plays the role of a $D_0$ -
dimensional effective cosmological constant $\Lambda_{eff}=\left. U_{eff}\right|_{min}$.
\item
 From eqs. \rf{3.20}
and \rf{3.22} follows that the compactified internal space should
have negative curvature.
\end{enumerate}
 The latter restriction agrees with the results of
\cite{CGHW,NSST}  because the negative value of the effective potential
in the minimum violates the null energy condition so that
 the stabilized internal space should be (compact) hyperbolic (see also \cite{demir,NSST}). We note that adding
to our non-linear model some kind of matter,
 satisfying the null energy condition,  can shift the effective
$D_0$-dimensional cosmological constant to
non-negative values and the internal space can acquire positive
curvature.

A further restriction on the model follows from eqs.  \rf{0.2}, \rf{2.5} and
\rf{3.19.1}. According to these equations the free parameters  $\alpha$ and $\Lambda
_D$, or $\alpha$ and $q$,
are strongly connected with the compactification radius $r_1$ of the extra dimensional factor space
$M_1$,
as well as with
the fundamental mass scale $M_{*(4+d_1)}$ and
the 4-dimensional Planck scale $M_{Pl(4)}$:
\be{3.25a}
\frac{2d_1}{D-2} \left. U \left[\phi_0(q),\alpha
\right]\right|_{extr}= R_1= -\frac{d_1(d_1-1)}{r_1^2}
\sim -
\left(\frac{M_{*(4+d_1)}}{M_{Pl(4)}}\right)^{4/d_1}M_{*(4+d_1)}^2\;
.
\ee
For fixed compactification radius $r_1<\infty$ the constraint
\rf{3.25a}
forbids the limit $\Lambda_D\to -0$, whereas $\alpha \to 0$ is
allowed.
This behavior is easily understood. According to \rf{1.18} the limit $\alpha \to 0$
describes the transition to a linear Einstein gravity model with
$D$-dimensional cosmological constant $\Lambda_D$. For $\alpha \to 0$ the mass of the $\phi$-field excitations
tends to infinity $m^2_{\phi}\to \infty$ (see eq. \rf{3.24.2} below) and the field itself becomes frozen at
the minimum position $\phi_0(\alpha\to 0)\to 0$ of the potential $U(\phi)$
\be{3.25a2}
\left.U[\alpha\to 0]\right|_{extr}\to \Lambda_D\; , \quad
\partial^2_{\phi \phi}\left.U\right|_{extr}\to \infty  .
\ee
The resulting $D$-dimensional space-time has constant scalar
curvature $\bar R = R = 2D\Lambda_D/(D-2)$ and a stabilization of
internal spaces in such models is possible \cite{GZ1} for  $\Lambda_D < 0$ and $R_i <
0$.

In contrast,
the  transition $\Lambda_D\to -0$ necessarily implies $\left. U(\phi) \right|_{extr} \to -0, \ R_1\to -0$
which is connected with a
decompactification $r_1\to \infty$ of the extra dimensions according
to \rf{3.25a}.
From the derivatives \rf{3.17.1} - \rf{3.17.3} of the effective potential at the extremum
position $(\varphi_{extr}=0,\ \phi_0)$ and
\be{3.25b}
\partial_{\varphi}^n\left.U_{eff}\right|_{extr}=-2^{n-1}\left[\frac{D-2}{d_1(D_0-2)}\right]^{n/2}R_1
+2^n\left[\frac{d_1}{(D-2)(D_0-2)}\right]^{n/2}\left.U\right|_{extr}
\ee
we read off that in the limit $\Lambda_D\to -0$ the potential becomes flat with
respect to
$\varphi $:\ $\partial_{\varphi}^nU_{eff}\to 0$, whereas it remains well-behaved with respect to $\phi$:
\be{3.25c}
\left.\partial^2_{\phi
\phi}U_{eff}\right|_{extr}\longrightarrow\frac{D-2}{4\alpha(D-1)}>0\;
.
\ee
This is due to $x_0(\Lambda_D\to 0)\to 1$ and eq. \rf{3.13}.
The potential $U_{eff}(\varphi,\phi)$ itself coincides in this case
with the effective potential of a model with Ricci-flat factor
space $M_1$
\be{3.25d}
U_{eff}(\varphi,\phi)=e^{2\varphi\sqrt{\frac{d_1}{(D-2)(D_0-2)}}}U(\phi)\;
,
\ee
what is known to have no stabilized extra dimensions. A
stabilization could be achieved, e.g., by accounting for additional
matter fields \cite{GZ1,GZ,GZ(PRD2),GMZ}.

Finally, let us turn to the
 masses
 of the excitation fields $\varphi$ and $\phi$ near the
minimum of $U_{eff}$. These masses are defined by the relations
\ba{3.24}
m^2_{\varphi }& = &
\left. \partial^2_{\varphi \varphi } U_{eff} \right|_{min} = -\frac{4}{D-2}
\left. U(\phi) \right|_{min} = -\frac{2}{d_1}\; R_1\; , \label{3.24.1}\\
m^2_{\phi }& = &
\left. \partial^2_{\phi \phi}U_{eff}\right|_{min} =
\partial^2_{\phi \phi}\left. U (\phi)\right|_{min}
= \frac{1}{4\alpha}\frac{1}{D-1} x_0^{-\frac{2}{D-2}}
\left[(D-4)x_0 +2\right] \; .\label{3.24.2} \ea
In the decompactification limit $r_1\to \infty, \  \Lambda_D\to -0, \ R_1\to
-0$ the mass of the gravexciton vanishes $m^2_{\varphi}\to
0$, whereas the mass of the $\phi-$field remains non-zero $m^2_{\phi }\to
(D-2)/(4\alpha(D-1))>0$.
For fixed compactification scale $r_1$ the constraint \rf{3.19.1} and its implication
\be{3.26} \frac{1}{4\alpha} = \frac{D-2}{d_1}x_0^{\frac{D}{D-2}}[
(x_0-1)^2 + q ]^{-1}\, R_1 \ee
can be used to express eq. \rf{3.24.2}
in terms of $x_0$ and $R_1$
\be{3.27} m_{\phi}^{2} =
\frac{D-2}{(D-1)d_1}\frac{x_0[(D-4)x_0+2]}{(x_0-1)^2 +q}\; R_1 \,
. \ee
This means that in an ADD scenario, where relation \rf{3.25a}
necessarily holds,
the basic mass scale of the excitations
 $\varphi$ and
$\phi$ is defined by the fundamental mass scale $M_{*(4+d_1)}$ and
the 4-dimensional Planck scale $M_{Pl(4)}$
 \be{3.27b}
 m_{\varphi,\phi}^2 \sim  R_1
 \sim r_1^{-2}\sim \left(\frac{M_{*(4+d_1)}}{M_{Pl(4)}}\right)^{4/d_1}M_{*(4+d_1)}^2.\ee

\section{\label{conclu}Conclusions and discussion}


In the present paper we investigated multidimensional
gravitational models with a non-Einsteinian form of the action. In
particular, we assumed that the action is an arbitrary smooth
function of the scalar curvature $f(R)$. For such models, we
concentrated on the problem of extra dimension stabilization in
the case of factorizable geometry. To perform such analysis, we
reduced the pure non-linear gravitational  model to a linear one
with an additional self-interacting scalar
field. The factorization of the geometry allowed for a dimensional
reduction of the considered models and to obtain an effective
4--dimensional model with additional minimally coupled scalar fields in the Einstein frame. These
fields describe conformal excitations of the internal space scale
factors. A detailed stability analysis was carried out for a model
with quadratic curvature term: $f(R) = R + \alpha R^2 -2\Lambda
_D$. It was shown that a stabilization is only possible for the
parameter range $-1 < 8\alpha \Lambda _D <0 $.

This necessarily implies that the extra dimensions are stabilized
if the compact internal spaces $M_i, \ \ i=1,\ldots ,n$ have
negative constant curvatures. More precisely, these spaces have a
quotient structure $M_i=H^{d_i}/\Gamma_i$, where $H^{d_i}$ and
$\Gamma_i$ are hyperbolic spaces and their discrete isometry groups,
respectively. In this case, the 4--dimensional cosmological
constant (which corresponds to the minimum of the effective
4--dimensional potential) is also negative. As a consequence, the
homogeneous and isotropic external $(D_0=4)$-dimensional space is
asymptotically $\mbox{AdS}_{D_0}$. As the extra dimensional scale
factors approach their stability position the bulk space-time
curvature asymptotically (dynamically) tends to a negative
constant value. Let us note that this would allow e.g. for a
spontaneous compactification scenario along the lines\footnote{An explicit
 generalized deSitter solution for a similar stabilized warped product space was
obtained in \cite{ds}. The warped product consisted of a
Ricci-flat or $R\times S^3$
 external space-time and
 Einstein spaces with positive constant scalar
curvatures as internal spaces.}
\be{4.1} \mbox{AdS}_D\longrightarrow \mbox{AdS}_{D_0}\times
H^{d_1}/\Gamma_1 \times \dots \times H^{d_n}/\Gamma_n \; .\ee

We further found that the compactification scale completely
defines the effective cosmological constant and the mass of the
internal scale factor excitations (gravexcitons) near the minimum
position.
It is also strongly connected with the parameters $\alpha$ and
$\Lambda _D$ of the non-linear model (see eq. \rf{3.26}).
For example, in the limit $\Lambda _D\to 0$ the extra dimensions
necessarily decompactify ($r_1\to\infty$) and the effective potential $U_{eff}$ becomes indistinguishable
from an effective potential of a model with Ricci-flat factor space $M_1$.
The corresponding scale factor is then not stabilized and the
gravexciton becomes massless. In contrast to models with possible decompactification,
 ADD scenarios are characterized by a compactification scale which
 is fixed by
relations \rf{0.1},
 \rf{0.2} between the fundamental mass scales
$ M_{Pl(4)} $ and $M_{*(4+D^{\prime})} $.
   The same relations enforce in this case a constraint on the parameters  of
   the non-linear model.  In contrast with the masses of gravexcitons
$m_{\varphi}$ which are completely fixed by the compactification scale, the mass $m_{\phi}$ of the scalar field $\phi$
(which originates from the non-linearity of the starting model)
can still depend in a specific way on the  parameters $\alpha$ and $\Lambda _D$.

From a cosmological perspective, it is of interest to consider the
possibility of inflation for the 4--dimensional external
space-time within  our non-linear model. For a linear
multidimensional model with an arbitrary scalar field (inflaton),
such an analysis was carried out in \cite{GZ(PRD2)}. As described in
section \ref{setup}, our pure gravitational quadratic curvature action
\rf{1.18} can be mapped to a scenario linear in the curvature
with a rather specific self-interaction potential \rf{1.20} for
the scalar field $\phi$. This allows us to extend some of the
techniques of \cite{GZ(PRD2)} to our model.

It can be shown that there is a possibility for inflation to occur
if the scalar fields start to roll down from the region:
\be{4.19} |U(\phi )| \ge |\left.U(\phi)\right|_{min}| \gg |R_1|
e^{2\varphi \sqrt{\frac{D_0-2}{d_1(D-2)}}}\; , \ee
where the effective potential \rf{3.3} reads
\be{4.20} U_{eff} \approx e^{2\varphi
\sqrt{\frac{d_1}{(D-2)(D_0-2)}}} U(\phi )\; . \ee
If
\be{4.21} e^{\sqrt{\frac{D-2}{D-1}}\phi} \gg 1 \; ,\ee
and hence $U(\phi) \approx \frac{1}{8\alpha} e^{(2A-B)\phi} =
\frac{1}{8\alpha} \exp{\frac{D-4}{\sqrt{(D-2)(D-1)}}\phi}$, the
slow-roll parameters $\epsilon $ and $\eta _{1,2}$ (see paper
\cite{GZ(PRD2)}) read
\be{4.22} \epsilon \approx \eta_1 \approx \eta_2 \approx
\frac{2d_1}{(D-2)(D_0-2)} + \frac{(D-4)^2}{2(D-2)(D-1)}\; . \ee
For the dimensionality of our observable Universe $D_0=4$, these
parameters are restricted to the range
\be{4.23} \frac35 \le \epsilon ,\eta_1 ,\eta_2 \le 1\quad
\mbox{for}\quad 6 \le D \le 10\; . \ee
Thus, generally speaking, the slow-roll conditions for inflation
are satisfied in this region. The scalar field $\phi$ can act as
inflaton and drive the inflation of the external space. It is
clear that estimates \rf{4.23} point only to the possibility for
inflation to occur. For the considered model with negative effective
cosmological constant inflation is not
successfully completed \cite{stein2} if the reheating
due to the decay of the $\phi -$field excitations and gravexcitons is not sufficient for a transition
to the radiation dominated era. In any case, for scenarios with
successful transition or without, the external space
has a turning point at its maximal scale factor where
the evolution changes from expansion to contraction\footnote{A discussion
of this effect can be found in \cite{GZ(PRD2)} and the recent
paper \cite{FFKL}.}. Obviously, for such models the negative effective cosmological
constant forbids a late time acceleration of the Universe as indicated by
recent observational data. In order to cure this problem, the model
should be generalized, e.g., by inclusion of additional form fields
\cite{GMZ} or other matter fields.

\vspace{.4cm}

\mbox{} \\ {\bf Acknowledgments}\\
 U.G. and A.Z. thank the
Department of Physics of the University of Beira Interior for kind
hospitality during the preparation of this paper. The work of A.Z.
was supported by a BCC grant from CENTRA--IST  and partly
supported by the programme SCOPES (Scientific co-operation between
Eastern Europe and Switzerland) of the Swiss National Science
Foundation, project No. 7SUPJ062239. U.G. acknowledges
support from DFG grant KON/1344/2001/GU/522. Additionally, this research work
was partially supported by the grants POCTI/32327/P/FIS/2000,
CERN/P/FIS/43717/2001 and CERN/P/FIS/43737/2001.



\begin{thebibliography}{99}
\bibitem{pol-wit} M.B. Green, J.H. Schwarz and E. Witten,
Superstring theory, Cambridge: Cambridge University Press, 1987;
J. Polchinski, String theory, Cambridge: Cambridge University
Press, 1998.
\bibitem{sub-mill1}
N. Arkani-Hamed, S. Dimopoulos and G. Dvali, Phys. Lett. B429,
(1998), 263 - 272, hep-ph/9803315.
\bibitem{sub-mill1a}
I. Antoniadis, N. Arkani-Hamed, S. Dimopoulos and G. Dvali, Phys.
Lett. B436, (1998), 257 - 263,  hep-ph/9804398.
\bibitem{sub-mill2}
N. Arkani-Hamed, S. Dimopoulos and G.J. March-Russell, Phys. Rev.
D63, (2001), 064020, hep-th/9809124.
\bibitem{experiment1}C.D. Hoyle et al, Phys. Rev. Let. 86, (2001), 1418,
hep-ph/0011014; G. Dvali, G. Gabadadze, X. Hou and E. Sefusatti,
{\it See-saw modification of gravity}, hep-th/0111266.
\bibitem{RS}
L. Randall and R. Sundrum, Phys. Rev. Let. 83, (1999), 3370,
hep-ph/9905221; ibid. 83, (1999), 4690, hep-th/9906064.
\bibitem{WCOS}
L. Wang, R.R. Caldwell, J.P. Ostriker and P.J. Steinhardt,
Astrophys. J. 530, (2000), 17 - 35, astro-ph/9901388.
\bibitem{d2}
T. Banks, M. Dine and A.E. Nelson, JHEP 9906, (1999), 014,
hep-th/9903019.
\bibitem{sub-mill3}
N. Arkani-Hamed, S. Dimopoulos, N. Kaloper and J. March-Russell,
Nucl. Phys. B567, (2000), 189 - 228, hep-ph/9903224.
\bibitem{CGHW} S.M. Carroll, J. Geddes, M.B. Hoffman and R.M.
Wald, {\it Classical stabilization of homogeneous extra dimensions},
Phys. Rev. D, to appear,
hep-th/0110149.
\bibitem{Geddes} J. Geddes, Phys. Rev. D65, (2002), 104015,
gr-qc/0112026.
\bibitem{demir}D.A. Demir and M. Shifman, Phys. Rev. D65, (2002),
104002, hep-ph/0112090.
\bibitem{NSST} S. Nasri, P.J. Silva, G.D. Starkman and M. Trodden,
{\it Radion stabilization in compact hyperbolic extra dimensions},
hep-th/0201063.
\bibitem{PS} L. Perivolaropoulos and C. Sourdis, {\it Cosmological
effect of radion oscillations}, hep-ph/0204155.
\bibitem{GZ1}  U. G\"unther and A. Zhuk, Phys. Rev. D56, (1997), 6391 - 6402,
gr-qc/9706050.
\bibitem{GZ}
U. G\"unther and A. Zhuk,  {\it Stable compactification and
gravitational excitons from extra di\-men\-sions}, (Proc. Workshop
''Modern Modified Theories of Gravitation and Cosmology'', Beer
Sheva, Israel, June 29 - 30, 1997), Hadronic Journal 21, (1998),
279 - 318, gr-qc/9710086;\\
U. G\"unther, S. Kriskiv and A. Zhuk, Gravitation \&
 Cosmology 4, (1998), 1 -16, gr-qc/9801013;\\
U. G\"unther and A. Zhuk, Class. Quant. Grav. 15,
(1998), 2025 - 2035, gr-qc/9804018.
\bibitem{GZ(PRD2)}
U. G\"unther and A. Zhuk, Phys. Rev. D61, (2000), 124001,
hep-ph/0002009.
\bibitem{CGR}
C. Charmousis, R. Gregory and V.A. Rubakov, Phys. Rev. D62,
(2000), 067505, \, hep-th/9912160.
\bibitem{Csaki1}C. Cs\'aki, M.L. Graesser, L. Randall and J.
Terning, Phys. Rev. D62, (2000), 045015,
    hep-ph/9911406.
\bibitem{Csaki2}C. Cs\'aki, M.L. Graesser and G.D. Kribs, Phys. Rev. D63, (2001), 065002,
    hep-th/0008151.
\bibitem{MBLSZ}M. Bouhmadi-Lopez and A. Zhuk,
Phys. Rev. D65, (2002), 044009,
    hep-th/0107227.
\bibitem{BDL}
P. Bin\'etruy, C. Deffayet and D. Langlois, Nucl. Phys. B615,
(2001), 219 - 236,\, hep-th/0101234.
\bibitem{CF}
Z. Chacko and P.J. Fox, Phys. Rev. D64, (2001), 024015, \,
hep-th/0102023.
\bibitem{Kerner} R. Kerner, GRG 14, (1982), 453 - 469;\\
J.D. Barrow and A.C. Ottewill, J. Phys. A 10, (1983), 2757 - 2776;\\
J.P. Duruisseu and R. Kerner, GRG 15, (1983), 797 - 807;\\
B. Whitt, Phys. Lett. B145, (1984), 176 - 178;\\
J.D. Barrow and S. Cotsakis, Phys. Lett. B214, (1988), 515 -
518;\\
K. Maeda, J.A. Stein--Schabes and T. Futamase, Phys. Rev. D39,
(1989), 2848 - 2853;\\
G. Magnano and L.M. Sokolowski, Phys. Rev. D50, (1994), 5039 -
5059, gr-qc/9312008;\\
D. Wands, CQG, 11 (1994) 269-279, gr-qc/9307034;\\
J. Ellis, N. Kaloper, K.A. Olive and J. Yokoyama, Phys. Rev. D59,
(1999), 103503, hep-ph/9807482.
\bibitem{IMZ}  V.D. Ivashchuk, V.N. Melnikov and A.I. Zhuk, Nuovo Cimento
B104,(1989), 575 - 582.
\bibitem{RZ}  M. Rainer and A. Zhuk, Phys. Rev. D54, (1996), 6186 - 6192,
(gr-qc/9608020).
\bibitem{GMZ}U. G\"unther, P. Moniz and A. Zhuk, {\it Non-linear multidimensional cosmological models with form
fields}, (in preparation).
\bibitem{ds}  A. Zhuk, Astron.
Nachr. 316, (1995), 269 - 274, gr-qc/0205116.
\bibitem{stein2}T. Banks, M. Berkooz and P.J. Steinhardt, Phys. Rev. D52, (1995), 705 - 716,
 hep-th/9501053.
\bibitem{FFKL}G. Felder, A. Frolov, L. Kofman and A. Linde, {\it Cosmology with negative
potentials}, Phys. Rev. D, to appear,  hep-th/0202017.

\end{thebibliography}
\end{document}